\pdfoutput=1 
\documentclass[%
reprint, twocolumn,
 amsmath,amssymb,
 aps,
]{revtex4}
\usepackage{graphicx}
\usepackage{dcolumn}
\usepackage{bm}
\usepackage{appendix}
\usepackage{pdfpages}

\begin{document}


\title{Translation of Werner Heisenberg's Paper "Quantum-Theoretical Reinterpretation of Kinematic and Mechanical Relations" Zeitschrift für Physik 33, 1925, 879-893 into Czech language}
\thanks{English translation of the original title {\it \"{U}ber quantentheoretische Umdeutung kinematischer und mechanischer Beziehungen} taken from B. L. Waerden, ed., {\it Sources of Quantum Mechanics}, Dover Publications, Inc., New York, 1968 }%

\author{Tom\'{a}\v{s} Man\v{c}al}
\affiliation{%
 Faculty of Mathematics and Physics, Charles University,  Ke Karlovu 5, CZ-121 16 Prague, Czech Republic
}%

\begin{abstract}
To facilitate teaching of quantum mechanics on undergraduate and even advanced secondary school levels, we present an unabridged translation of the original German text of the famous Werner Heisenberg's breakthrough paper of 1925 into Czech language. While the paper introduces fundamentally new concepts into mechanics, and it contains a relatively large number of equations, its conceptual side is surprisingly lightweight when it comes to the use of advanced mathematics. The discussions contained in its Section 1 allow students of quantum mechanics to contemplate the reasons for the inadequacy of classical kinematic concepts in the micro-world. Understanding of the kinematic transition from real valued functions to complex valued "matrices", which is the central point of the paper, can be reached with relatively elementary mathematics. Translations of the Heisenberg's work into languages other than English provide a unique opportunity to introduce the great scientific concepts of quantum mechanics first hand to a broader audience of students who have yet to reach proficiency in the current scientific {\it{lingua franca}}. The present translation is equipped with footnotes explaining the notation, identifying several typos and pointing out sudden changes in notation in the original paper.

\end{abstract}

\maketitle


\section{\label{sec:level1}Introduction}

In his famous paper \cite{Be-1925}, {\it \"{U}ber quantentheoretische Umdeutung kinematischer und mechanischer Beziehungen}, often referred to only as {\it Umdeutung},  Werner Heisenberg formulated the first version of modern quantum mechanics. This formulation later merged with the equivalent, but formally different formulation of Erwin Schrödinger, developed around the same time. Historically, the paper was quickly superseded by a mathematically much more concise matrix mechanics of Born and Jordan (and eventually Heisenberg himself) \cite{Born1925, Born1926}, and it might therefore seem of a historical value  only. Quantum mechanics as we know and teach it today is barely recognizable in the  {\it Umdeutung} paper, and Heisenberg's train of thoughts was deemed incomprehensible even by some of the most accomplished theoretical physicists of our days \cite{Weinberg1992}.
Nevertheless, close examination of the paper in the context of works which preceded it \cite{Blum2017}, and successful attempts to reconstruct Heisenberg's calculations (whose details are largely missing in the paper itself) \cite{Aitchison2004}, seem to reveal a rich source of material for introductory discussion of quantum mechanics \cite{DiMauro2021}.

In his introductory paragraphs, Heisenberg provides some justification for his dismissal of classical kinematics in the quantum realm. As he reveals already in his brief abstract, he strives to construct a theory based on observable quantities only. Such a motivation is very likely insufficient to rigorously justify Heisenberg's steps, as noted by philosophers and historians of science, see \cite{Beller1999, Wuthrich2016}, and as confirmed by the fact that the later successful quantum theory operates with another hardly observable quantity, the state vector. Heisenberg's motivation was even criticised by Einstein as being "quite wrong" approach to theoretical science "on principle" (see \cite{Heisenberg1971}, p. 63, or quoted in \cite{Wuthrich2016} in a different translation). Students would therefore not be well served, if Heisenberg's introductory paragraphs were presented to them as representative of some rigorous scientific methodology. Nevertheless, the technical side of Heisenberg's discussion undoubtedly sets the stage for his crucial proposition, namely, to replace classical kinematics based on the real functions, with a new kinematic concept of an ensemble of complex quantities (later recognized as matrices by Born \cite{Born1925}). Heisenberg, in his introductory paragraphs, points out the failure of the electron orbit concept in the theoretical attempts of the so-called Old Quantum Theory (OQT) to understand multi-electron atoms. As he thinks in terms of observable quantities, Heisenberg always stays close to the theory of light-matter interaction. He sees the failures of the OQT, which is  essentially just classical physics augmented by some additional rules and constrains, as a failure of the classical kinematics - classical description of light absorption and emission is at odds with empirical laws deduced from observations of atomic phenomena. Heisenberg's brief discussion of the elementary classical theory of radiation serves one important goal, namely, to show how the electron motion, represented by a time-dependent function of position and its time derivatives, enters the expressions for the radiated fields. As the classical quantities enter the expression for fields in various powers, the central questions concerning the new kinematics is posed by Heisenberg the following way: {\it If instead of a classical quantity $x(t)$ we have a quantum-theoretical quantity, what quantum-theoretical quantity will appear} [in the theory of radiation] {\it in place of} $x^2(t)$ \cite{BLvanderWaerden}.

To answer this question, Heisenberg compares the main properties of radiation frequencies as predicted by the classical theory with those expected from the (future) quantum theory. He identifies a new rule of combination (multiplication) of oscillatory components of the electron's periodic motion, which ensures that only those frequencies, which satisfy the (quantum) Ritz combination principle, ever appear in the quantum theory of radiation (in an arbitrary order of perturbation theory). His new rules allow him to demonstrate validity of the new kinematics by solving several one-dimensional problems, such as the harmonic and anharmonic oscillators.

While the application part of the paper is very technical, its conceptual part, in which the new kinematics is motivated and constructed, can be turned into an accessible narrative motivating the highly non-trivial step of introducing operators to the mechanical theory of atomic phenomena. All this with secondary school mathematics only.

The present paper provides an unabridged translation of the {\it Umdeutung} paper into Czech language, and it suggests a narrative through which sections of the paper can be used to explain fundamental concepts of quantum mechanics to undergraduate and advanced secondary school students, as well as to interested lay public. In Section \ref{sec:usage} we propose the outline of such an exposition of quantum mechanics to the students. Section \ref{sec:trans} is concerned with a brief description of the translation, and with the discussion of the footnote explaining difficult or ambiguous places of Heisenberg's paper. The translation itself is present in the Appendix.

\section{\label{sec:usage}Possible usage of the {\it Umdeutung} paper in education}
Original scientific papers provide a unique perspective on the development and practices of science. This perspective is useful to both students and interested lay readership \cite{lightman2006}. In the age of open access publication, the latest advances in science become accessible to interested readers, provided that they are proficient in English. Historical development of quantum theory is one of the last major scientific revolutions, which are originally recorded mainly in a language other than English. However, the importance of the first hand access to the founding papers of quantum mechanics for the present and future generation of historians and educators is such that the most important original works were translated into English relatively early on \cite{BLvanderWaerden}.

Using Heisenberg's {\it Umdeutung} paper in physics education is certainly not a new idea \cite{DiMauro2021}. In this section, we will describe a possible narrative motivating, on the basis of the {\it Umdeutung} paper, the necessity to replace the real valued functions of classical mechanics with some new mathematical objects more suitable for the description of atomic phenomena. The narrative describes tentative thoughts of a theoretical physicist, who, facing unsolved problems in atomic physics, searches for new mathematics satisfying the known properties of atomic systems. Although we refer to Werner Heisenberg in the sections below, we do not assume anything about his actual historical thoughts and motivations. We only create a plausible narrative based on his text to allow students, with the assistance of the teachers, to rediscover the quantum "multiplication" rules for themselves.

\paragraph{Classical Periodic Motion.} Before they venture into the quantum physics, the students need to get acquainted with the properties of the classical periodic motion. Just as in Heisenberg's paper, the discussion can be conveniently limited to the motion of a particle in a one dimensional potential only. Starting with a particle at rest at a certain displacement from the bottom of the potential, it is possible to visually demonstrate to the students (using graphs or computer animation) that the particle's motion can be constructed by adding together cosines of some basic frequency $\omega$, and its integer multiples (the higher harmonics). The frequency is obtained directly from the period $T$ of the classical motion as $
    \omega = \frac{2\pi }{T}.
$
The position $x(t)$ of the particle as a function of time can be then written down as
\begin{equation}
\label{eq:fourier}
    x(t)=\sum_{n=0}^{\infty} b_n \cos(n\omega t)=\sum_{n=-\infty}^{\infty} a_n e^{i n\omega t},
\end{equation}
where $a_n = b_n / 2$ for $n>0$ and $a_n = b_{-n} / 2$ for $n<0$. All this can be explained without any specific reference to Fourier series. At this point it needs to be discussed with the students that classical theory predicts that electromagnetic waves originating from the particle's periodic motions will contain components with frequencies $n\omega$, for $n=1,2,\dots$. This is a fact well known to Heisenberg, and it is implicitely used in his introductory sections.

\paragraph{Frequencies of the classical and quantum periodic motion.} The second step in the exposition of Heisenberg's ideas is the discussion of Bohr's version of the Ritz combination principle. This principle states that the frequencies of light emitted from quantum systems, such as atoms, correspond to differences in energy between stationary states of the systems:
\begin{equation}
\label{eq:ritz}
    \omega_{nm} = \frac{\varepsilon_n-\varepsilon_n}{\hbar}.
\end{equation}
Here, one should point out and discuss the crucial difference in behaviour of the classical and quantum frequencies. The classical frequencies are all integer multiples of a certain fundamental frequency $\omega$. As a consequence, addition of any two allowed frequencies leads to an allowed frequency:
\begin{equation}
    \label{eq:class_comp}
    n\omega + m\omega = (n+m)\omega.
\end{equation}
On the contrary, the quantum frequencies as differences between pairs from a given set of real numbers (system's energies) cannot be arbitrarily combined to return an allowed quantum frequency. Adding two frequencies
\begin{equation}
    \label{eq:quant_comp}
    \omega_{nm} + \omega_{kl} = \frac{\epsilon_n - \epsilon_m + \epsilon_k - \epsilon_l}{\hbar},
\end{equation}
does not yield an allowed frequency in the form of Eq. (\ref{eq:ritz}) unless two of the four energies cancel. This is only possible if $m=k$ or and $n=l$ (with the exception of accidental degeneracy in the spectrum). Why is this important will become clear in the next step of the exposition.

\paragraph{Classical kinematics} Classical coordinate $x(t)$ of the periodic motion and its time derivatives enter the expressions for the radiation in various powers (see \cite{Be-1925}, Section 1). Heisenberg's kinematic questions is: how to calculate the oscillatory components of quantities such as $x(t)$ and $y(t)$, when the corresponding coefficients in Eq. (\ref{eq:fourier}) for $x(t)$ and $y(t)$ are known? Classical rule for combining the coefficients is straightforwardly obtained from ordinary rules of multiplication. The multiplication of $x(t)y(t)$ amounts to multiplication of two polynomials, in which all terms of $x(t)$ multiply all terms of $y(t)$, e.g.
\begin{equation}
\label{eq:coeff_1}
    a_n e^{in\omega t}b_m e^{im\omega t} = a_n b_m e^{i(n+m)\omega t}.
\end{equation}
If we are to calculate a coefficient of the frequency $k\omega$, where $k$ is an integer, in the expansion of $x(t)y(t)$, we collect all coefficients, Eq. (\ref{eq:coeff_1}), such that $n+m=k$. This gives
\begin{equation}
    \label{eq:rule_class}
    c_k = \sum_n a_n b_{k-n}.
\end{equation}
Here, we removed the coefficient $e^{ik\omega t}$ from the both sides of Eq. (\ref{eq:rule_class}).
Because $(n+m)\omega$ is always a valid frequency of the motion, there is no conflict between the classical rules of frequency composition, Eq. (\ref{eq:class_comp}), and the classical description of motion by real valued functions.

\paragraph{Quantum kinematics}
As far as we can deduce from Heisenberg's paper, in the search for a new quantum kinematics, his main idea is to make sure that the quantum frequency composition rules, Eq. (\ref{eq:quant_comp}), is automatically satisfied in all operations with quantum quantities. This is in line with the function that kinematics usually has in physics. The notion of electrons orbit (or position, path) $x(t)$ has failed to deliver meaningful results in the OQT. Heisenberg therefore frees himself from $x(t)$, Eq. (\ref{eq:fourier}), as a sum of oscillatory coefficients, and concentrates on the oscillatory components themselves. They, as an ensemble of quantities, should now carry the physics of the periodic motion. Quantum description of periodic motion is thus based on oscillatory terms
\begin{equation}
\label{eq:qcoeffs}
    a_{nm}e^{i\omega_{nm}t},
\end{equation}
where the coefficients $a_{nm}$ now carry two indices, to reflect the double-index nature of the quantum frequencies. Just as the classical coefficients $a_n$ enter the expression for the oscillatory component of the classical radiation field (through expressions for $\dot{x}(t)$, $\ddot{x}(t)$ and so on), Heisenberg expects that terms such as those in Eq. (\ref{eq:qcoeffs}) will enter corresponding quantum expressions for the radiation field.

Let us now represent the quantity $x(t)$ by coefficients $a_{nm}e^{i\omega_{nm}t}$, and the quantity $y(t)$ by coefficients $b_{nm}e^{i\omega_{nm}t}$. In order for the quantum frequency combination rule to be satisfied in operations with these coefficients, one cannot let all members of the ensemble representing $x(t)$ multiply all members of the ensemble representing $y(t)$ to form representation of $x(t)y(t)$. Only those coefficient pairs which return valid frequencies can meet in a multiplication, i.e., only terms such as
\begin{equation}
    a_{nm}b_{mk} e^{i\omega_{nm}t}e^{i\omega_{mk}t}=a_{nm}b_{mk}e^{i\omega_{nk}t},
\end{equation}
are allowed. Collecting then all elements resulting in a given frequency $\omega_{nk}$ we obtain
\begin{equation}
\label{eq:multiplic_rule}
    c_{nk} = \sum_{m} a_{nm}b_{mk},
\end{equation}
as the coefficients representing $x(t)y(t)$.
In Eq. (\ref{eq:multiplic_rule}) we removed the phase factor $e^{i\omega_{nk}t}$ common to both sides. Importantly, in Eq. (\ref{eq:multiplic_rule}) one can readily recognize matrix multiplication.

It is now irrelevant whether the coefficients representing the quantum analogue of a classical quantity are organized into matrices or not. The coefficients represent possible contributions to the radiation on the given frequency $\omega_{nk}$, and the rule of their combination, Eq. (\ref{eq:multiplic_rule}), ensures that any function of the quantum analog of $x(t)$ continues to deliver only frequencies allowed by the Ritz combination principle, Eq. (\ref{eq:ritz}).

The students can now easily demonstrate on an example of small finite matrices \cite{DiMauro2021}, that the matrix multiplication is not generally commutative, i.e. that for the new mathematical objects $\hat{x}(t)$ and $\hat{y}(t)$, the collections of oscillatory coefficients,  we have $\hat{x}(t)\hat{y}(t)\neq\hat{y}(t)\hat{x}(t)$.
Such a behaviour of the newly constructed quantities, which after all still represent the motion of a quantum particle, lends further support to the idea that the classical path and position concepts are inadequate in the atomic realm. In order to fulfill the requirements for quantum light frequencies, newly constructed quantities necessarily lose the basic properties of real functions.

Besides the introduction of the concept of matrices (operators), Heisenberg's paper also formulates the quantization condition of OQT in a form, which was later rewritten using the famous commutator of coordinate and momentum operators. This fact was only recognized later by Born and Jordan \cite{Born1925}. Another possibly valuable point of the paper, which gets often completely overlooked in teaching quantum mechanics, is the use of correspondence principle, i.e. the discussion of the classical limits of the quantum mechanics, which certainly played an important role in its invention. This feature of the {\it Umdeutung} paper, as well as the several example calculations which Heisenberg presents, are beyond the scope of this short paper, although even here Heisenberg's paper might be of interest to quantum mechanics students and educators.

\section{\label{sec:trans} The Translation into Czech language}

To enable an easy access to the complete text of the {\it Umdeutung} paper to advanced undergraduate or secondary school students of the Czech educational system, who do not speak German or are not yet proficient enough in English (where a translation is available), we performed a complete translation of its original German text into Czech language.

The translation follows the original text as faithfully as possible, including an attempt to keep the structure of sentences (as far as the differences between the source and target languages allow), and keeping the original mathematical notation (with a single exception described in the footnotes). The Czech text is typeset in such a way that the division into pages is kept as close to the original as possible, and the two texts can be read side by side. Original footnotes (Arabic numerals) of the paper are kept and also translated, and a system of new footnotes (lowercase letters) is added to comment the paper or details of the translation.

The original German text of the {\it Umdeutung} paper contains errors in Eqs. (16) and two equations above Eq. (17). We do not correct the errors in the translation, but we point them out and explain in the footnotes. In the application part of the paper, its Section 3, Heisenberg switches from using $a(n)$ for the coefficients of complex exponentials in the Fourier series, into denoting the coefficients of cosines by the same expression. This results in a seemingly missing factor of $4$ in the last equation before Eq. (20) on page 888 of the German text. The reader is alerted to this sudden switch of notation in the new footnotes.

The translation was compared to the available English translation from Ref. \cite{BLvanderWaerden}. Several more difficult places of the present translation were harmonized with the English version. The differences in the meaning between the two translation are minimal.

\section{Conclusions}
Recent progress in quantum technologies, and the media attention these technologies attract nowadays, bring quantum mechanics to the forefront of interest of ever younger students. Young people at the undergraduate level or even advanced secondary school students seeking insight into quantum mechanics face a likely dilemma of either succumbing to the heavy load of advanced mathematics, or staying on a surface with only "second hand" descriptive accounts of the subject. The possibility to use, first hand, sections of the original discovery papers to introduce quantum mechanics seems remote at the first glance. However, with proper set of support materials, and a support from teachers, Werner Heisenberg's {\it Umdeutung} paper can be used for just this purpose. Although English occupies an uncontested position of the scientific {\it lingua franca}, with all the benefits which this situation brings, it cannot be expected in foreseeable future (at least in Europe) that all young students interested in physics will be sufficiently proficient in English at the time that they first reach sufficient knowledge of mathematics to understand the core of Heisenberg's propositions. Translations into native languages may serve to overcome the language barrier which otherwise stand between a young student and her first hand experience of scientific literature. Although some of the founding papers of quantum mechanics might be written in a somewhat archaic style, they communicate a unique spirit of discovery, which invites students to take on the adventures of science.

The present translation represents the first part of a set of teaching materials to support a first mathematical exposition of quantum mechanics to undergraduate university students, advanced secondary school students, or interested members of public.


\begin{thebibliography}{12}
\expandafter\ifx\csname natexlab\endcsname\relax\def\natexlab#1{#1}\fi
\expandafter\ifx\csname bibnamefont\endcsname\relax
  \def\bibnamefont#1{#1}\fi
\expandafter\ifx\csname bibfnamefont\endcsname\relax
  \def\bibfnamefont#1{#1}\fi
\expandafter\ifx\csname citenamefont\endcsname\relax
  \def\citenamefont#1{#1}\fi
\expandafter\ifx\csname url\endcsname\relax
  \def\url#1{\texttt{#1}}\fi
\expandafter\ifx\csname urlprefix\endcsname\relax\def\urlprefix{URL }\fi
\providecommand{\bibinfo}[2]{#2}
\providecommand{\eprint}[2][]{\url{#2}}

\bibitem[{\citenamefont{Heisenberg}(1925)}]{Be-1925}
\bibinfo{author}{\bibfnamefont{W.}~\bibnamefont{Heisenberg}},
  \bibinfo{journal}{Zeitschrift f{\"{u}}r Physik}
  \textbf{\bibinfo{volume}{33}}, \bibinfo{pages}{879} (\bibinfo{year}{1925}).

\bibitem[{\citenamefont{Born and Jordan}(1925)}]{Born1925}
\bibinfo{author}{\bibfnamefont{M.}~\bibnamefont{Born}} \bibnamefont{and}
  \bibinfo{author}{\bibfnamefont{P.}~\bibnamefont{Jordan}},
  \bibinfo{journal}{Zeitschrift f{\"{u}}r Physik}
  \textbf{\bibinfo{volume}{34}}, \bibinfo{pages}{858} (\bibinfo{year}{1925}).

\bibitem[{\citenamefont{Born et~al.}(1926)\citenamefont{Born, Heisenberg, and
  Jordan}}]{Born1926}
\bibinfo{author}{\bibfnamefont{M.}~\bibnamefont{Born}},
  \bibinfo{author}{\bibfnamefont{W.}~\bibnamefont{Heisenberg}},
  \bibnamefont{and} \bibinfo{author}{\bibfnamefont{P.}~\bibnamefont{Jordan}},
  \bibinfo{journal}{Zeitschrift f{\"{u}}r Physik}
  \textbf{\bibinfo{volume}{35}}, \bibinfo{pages}{557} (\bibinfo{year}{1926}).

\bibitem[{\citenamefont{Weinberg}(1992)}]{Weinberg1992}
\bibinfo{author}{\bibfnamefont{S.}~\bibnamefont{Weinberg}},
  \emph{\bibinfo{title}{{Dreams of a Final Theory}}}
  (\bibinfo{publisher}{Pantheon}, \bibinfo{address}{New York},
  \bibinfo{year}{1992}).

\bibitem[{\citenamefont{Blum et~al.}(2017)\citenamefont{Blum, J{\"{a}}hnert,
  Lehner, and Renn}}]{Blum2017}
\bibinfo{author}{\bibfnamefont{A.}~\bibnamefont{Blum}},
  \bibinfo{author}{\bibfnamefont{M.}~\bibnamefont{J{\"{a}}hnert}},
  \bibinfo{author}{\bibfnamefont{C.}~\bibnamefont{Lehner}}, \bibnamefont{and}
  \bibinfo{author}{\bibfnamefont{J.}~\bibnamefont{Renn}},
  \bibinfo{journal}{Studies in History and Philosophy of Science Part B -
  Studies in History and Philosophy of Modern Physics}
  \textbf{\bibinfo{volume}{60}}, \bibinfo{pages}{3} (\bibinfo{year}{2017}).

\bibitem[{\citenamefont{Aitchison et~al.}(2004)\citenamefont{Aitchison,
  MacManus, and Snyder}}]{Aitchison2004}
\bibinfo{author}{\bibfnamefont{I.~J.~R.} \bibnamefont{Aitchison}},
  \bibinfo{author}{\bibfnamefont{D.~A.} \bibnamefont{MacManus}},
  \bibnamefont{and} \bibinfo{author}{\bibfnamefont{T.~M.}
  \bibnamefont{Snyder}}, \bibinfo{journal}{American Journal of Physics}
  \textbf{\bibinfo{volume}{72}}, \bibinfo{pages}{1370} (\bibinfo{year}{2004}).

\bibitem[{\citenamefont{Di~Mauro and Naddeo}(2021)}]{DiMauro2021}
\bibinfo{author}{\bibfnamefont{M.}~\bibnamefont{Di~Mauro}} \bibnamefont{and}
  \bibinfo{author}{\bibfnamefont{A.}~\bibnamefont{Naddeo}},
  \bibinfo{journal}{Phys. Sci. Forum} \textbf{\bibinfo{volume}{2}},
  \bibinfo{pages}{8} (\bibinfo{year}{2021}).

\bibitem[{\citenamefont{Beller}(1999)}]{Beller1999}
\bibinfo{author}{\bibfnamefont{M.}~\bibnamefont{Beller}},
  \emph{\bibinfo{title}{{Quantum Dialog: the making of a revolution}}}
  (\bibinfo{publisher}{University of Chicago Press},
  \bibinfo{address}{Chicago}, \bibinfo{year}{1999}).

\bibitem[{\citenamefont{W{\"{u}}thrich}(2016)}]{Wuthrich2016}
\bibinfo{author}{\bibfnamefont{A.}~\bibnamefont{W{\"{u}}thrich}}, in
  \emph{\bibinfo{booktitle}{The Philosophy of Historical Case Studies}}, edited
  by \bibinfo{editor}{\bibfnamefont{T.}~\bibnamefont{Sauer}} \bibnamefont{and}
  \bibinfo{editor}{\bibfnamefont{R.}~\bibnamefont{Scholl}}
  (\bibinfo{publisher}{Springer}, \bibinfo{year}{2016}), pp.
  \bibinfo{pages}{285--296}.

\bibitem[{\citenamefont{Heisenberg}(1971)}]{Heisenberg1971}
\bibinfo{author}{\bibfnamefont{W.}~\bibnamefont{Heisenberg}},
  \emph{\bibinfo{title}{{Physics and Beyond: Encounters and Conversations}}}
  (\bibinfo{publisher}{Harper and Row}, \bibinfo{address}{New York},
  \bibinfo{year}{1971}).

\bibitem[{\citenamefont{van~der Waerden}(1968)}]{BLvanderWaerden}
\bibinfo{author}{\bibfnamefont{B.~L.} \bibnamefont{van~der Waerden}},
  \emph{\bibinfo{title}{{Sources of quantum mechanics}}}
  (\bibinfo{publisher}{Dover Publications}, \bibinfo{address}{New York},
  \bibinfo{year}{1968}).

\bibitem[{\citenamefont{Lightman}(2006)}]{lightman2006}
\bibinfo{author}{\bibfnamefont{A.}~\bibnamefont{Lightman}},
  \emph{\bibinfo{title}{{The Discoveries: Great Breakthroughs in 20th Century
  Science}}} (\bibinfo{publisher}{Vintage Books}, \bibinfo{address}{New York},
  \bibinfo{year}{2006}).

\end{thebibliography}

\onecolumngrid

\appendix
\section{Translation of the {\it Umdeutung} paper into Czech}


%


\section*{Supplementary material}

# O novém kvantově teoretickém významu kinematických a mechanických vztahů.[a]

Od **W. Heisenberga** v Göttingenu

(Došlo 29. července 1925)

V této práci se pokusíme získat základy pro kvantově-teoretickou mechaniku, která se zakládá výhradně na vztazích mezi principiálně pozorovatelnými veličinami.

Je známo, že lze proti formálním pravidlům, která jsou obecně používána v kvantové teorii k výpočtu pozorovaných veličin (např. energie vodíkového atomu), vznést závažnou námitku, že tato výpočetní pravidla obsahují jako svoji podstatnou složku vztahy mezi veličinami, které pravděpodobně nelze principiálně pozorovat (jako např. pozici, oběžnou dobu elektronu), že tedy taková pravidla zjevně postrádají názorný fyzikální základ, pokud se ovšem nechceme stále ještě upínat k naději, že by se tyto doposud nepozorovatelné veličiny snad později mohly stát experimentu přístupnými. Tuto naději by bylo možno považovat za oprávněnou, pokud by zmíněná pravidla byla použitelná vnitřně konzistentně[b] a na přesně vymezenou oblast kvantově mechanických problémů. Zkušenost ale ukazuje, že se těmto formálním pravidlům podrobuje pouze atom vodíku a jeho Starkův jev, že ale již v problému „zkřížených polí" (atom vodíku v elektrickém a magnetickém poli různých směrů) vystupují fundamentální obtíže, že reakce atomu na periodicky se měnící pole s jistotou nelze pomocí zmíněných pravidel popsat, a že se nakonec rozšíření kvantových pravidel na popis atomu s více elektrony ukázalo jako nemožné. Je běžné tato selhání kvantově teoretických pravidel, která jsou v zásadě charakterizována použitím klasické mechaniky, popisovat jako odchylku od klasické mechaniky. Takové označení lze stěží považovat za smysluplné, uvážíme-li, že již (všeobecně platné) E i n s t e i n o v y – B o h r o v y frekvenční podmínky představují tak úplné odmítnutí klasické mechaniky nebo lépe, z pohledu vlnové teorie, odmítnutí kinematiky[c], na které se tato mechanika zakládá, že už při nejjednodušších kvantově teoretických problémech nelze naprosto pomýšlet na platnost klasické mechaniky. Za tohoto stavu věcí se zdá být vhodnější, vzdát se jakékoliv naděje na pozorování

doposud nepozorovatelných veličin (jako poloha[d], oběžná doba elektronu), současně připustit, že částečný souhlas jmenovaných kvantových pravidel se zkušeností je více méně náhodný, a pokusit se vybudovat, analogicky ke klasické mechanice, kvantově teoretickou mechaniku, ve které se vyskytují jen vztahy mezi pozorovatelnými veličinami. Za nejdůležitější první předpoklady[e] takové kvantově teoretické mechaniky můžeme považovat vedle frekvenční podmínky Kramersovu disperzní teorii[1] a na ní stavějící práce[2][f]. V následujícím chceme ukázat některé nové kvantově mechanické vztahy a použít je k úplnému pojednání některých speciálních problémů. Omezíme se při tom na problémy s jedním stupněm volnosti.

§ 1. V klasické teorii je záření pohybujícího se elektronu (v radiační oblasti, tj. $\mathfrak{E} \sim \mathfrak{H} \sim \frac{1}{r}$) dáno nejen pomocí vztahů[g]:

$$\mathfrak{E} = \frac{e}{r^3 c^2}\bigl[r[r\dot{v}]\bigr],$$

$$\mathfrak{H} = \frac{e}{r^2 c^2}[\dot{v}\,r],$$

nýbrž v dalším přiblížení se vyskytují ještě členy, např. ve tvaru

$$\frac{e}{rc^3}\dot{v}v,$$

které můžeme označit jako „kvadrupólové záření", v ještě vyšších řádech přiblížení členy např. ve formě

$$\frac{e}{rc^4}\dot{v}v^2;$$

tímto způsobem lze přiblížení protáhnout libovolně daleko. (V předcházejícím označují: $\mathfrak{E}$, $\mathfrak{H}$ intenzity polí ve zkoumaném bodě, $e$ náboj elektronu, $r$ vzdálenost elektronu od zkoumaného bodu, $v$ rychlost elektronu.)

Je možné se ptát, jak by musely vypadat tyto vyšší členy v kvantové teorii. Jelikož v klasické teorii lze vyšší přiblížení jednoduše vypočítat, je-li znám pohyb elektronu, popř. jeho Fourierův rozklad, budeme něco podobného očekávat i

---

[1] H. v.Kramers, Nature **113**, 673, 1924
[2] M. Born, ZS. f. Phys. **26**, 3798, 1924. H. A. Kramers und W. Heisenberg, ZS. F. Phys. **31**, 681, 1925. M. Born und P. Jordan, ZS. F. Phys. (v tisku)

v kvantové teorii. Tato otázka nemá nic společného s elektrodynamikou, nýbrž je, a to se nám zdá obzvláště důležité, čistě kinematické povahy; můžeme ji v nejjednodušší formě položit následujícím způsobem: Nechť je dána kvantově teoretická veličina vystupující na místě klasické veličiny $x(t)$; jaká kvantově teoretická veličina vystupuje na místě veličiny $x^2(t)$?

Než budeme moci tuto otázku zodpovědět, musíme si připomenout, že v kvantové teorii není možné přiřadit elektronu bod v prostoru jako funkci času pomocí pozorovatelných veličin. Elektronu však můžeme také v kvantové teorii přiřadit nějaké vyzařování; toto záření bude předně popsáno frekvencemi, které vystupují jako funkce dvou proměnných, kvantově ve tvaru:

$$\nu(n, n-\alpha) = \frac{1}{h}\{W(n) - W(n-\alpha)\},$$

v klasické teorii ve formě:

$$\nu(n, \alpha) = \alpha \cdot \nu(n) = \alpha \frac{1}{h}\frac{dW}{dn}.$$

(Tady jsme přiřadili $n \cdot h = J$ jedné z kanonických konstant.)

Jako charakteristické pro srovnání klasické a kvantové teorie z pohledu frekvencí můžeme vypsat kombinační vztahy:

Klasicky:
$$\nu(n, \alpha) + \nu(n, \beta) = \nu(n, \alpha + \beta).$$

Kvantově teoreticky:
$$\nu(n, n-\alpha) + \nu(n-\alpha, n-\alpha-\beta) = \nu(n, n-\alpha-\beta)$$
popř.   $\nu(n-\beta, n-\alpha-\beta) + \nu(n, n-\beta) = \nu(n, n-\alpha-\beta).$

Mimo frekvencí jsou následně k popisu záření nutné amplitudy; amplitudy mohou být chápány jako komplexní vektory (se šesti nezávislými koeficienty) určující[h] polarizaci a fázi. Také ony jsou funkcemi stejných dvou proměnných $n$ a $\alpha$, takže odpovídající složka záření bude představována následujícím výrazem:

Kvantově teoreticky:
$$Re\{\mathfrak{A}(n, n-\alpha)e^{i\omega(n,n-\alpha)t}\}. \qquad (1)$$
Klasicky:
$$Re\{\mathfrak{A}_\alpha(n)e^{i\omega(n)\cdot\alpha t}\}. \qquad (2)$$

Na první pohled se může zdát, že fázi (obsažené v $\mathfrak{U}$) nepřísluší v kvantové teorii žádný fyzikální význam[i], neboť frekvence nejsou v kvantové teorii obecně souměřitelné[j] s jejich vyššími harmonickými[k]. Ihned však uvidíme, že i v kvantové teorii má fáze určitý význam analogický klasické teorii. Zabýváme-li se nyní nějakou veličinou $x(t)$ můžeme si ji představovat jako reprezentovanou nějakým souborem veličin tvaru

$$\mathfrak{U}_\alpha(n)e^{i\omega(n)\cdot\alpha t},$$

které sjednocené sumou nebo integrálem, podle toho jedná-li se o periodický pohyb nebo ne, představují $x(t)$:[1]

popř.
$$\left.\begin{array}{l} x(n,t) = \displaystyle\sum_{\alpha=-\infty}^{\infty} \mathfrak{U}_\alpha(n)e^{i\omega(n)\cdot\alpha t} \\ x(n,t) = \displaystyle\int_{-\infty}^{\infty} \mathfrak{U}_\alpha(n)e^{i\omega(n)\cdot\alpha t}d\alpha. \end{array}\right\} \quad (2a)$$

Takové sjednocení odpovídajících kvantově teoretických veličin se v důsledku rovnocennosti veličin $n$, $n-\alpha$ nezdá být možné bez jisté libovůle, a tudíž se nezdá smysluplné; nicméně soubor veličin

$$\mathfrak{U}(n,n-\alpha)e^{i\omega(n,n-\alpha)t}$$

lze dobře chápat jako reprezentaci veličiny $x(t)$ a lze tedy hledat odpověď na výše položenou otázku: Čím bude reprezentována veličina $x(t)^2$?

Klasicky zní odpověď zřejmě[m] takto:

$$\mathfrak{B}_\beta(n)e^{i\omega(n)\beta t} = \sum_{\alpha=-\infty}^{\infty} \mathfrak{U}_\alpha \mathfrak{U}_{\beta-\alpha} e^{i\omega(n)(\alpha+\beta-\alpha)t} \quad (3)$$

popř.
$$= \int_{-\infty}^{\infty} \mathfrak{U}_\alpha \mathfrak{U}_{\beta-\alpha} e^{i\omega(n)(\alpha+\beta-\alpha)t}\,d\alpha \quad (4)$$

takže pak

$$x(t)^2 = \sum_{\beta=-\infty}^{\infty} \mathfrak{B}_\beta(n)e^{i\omega(n)\beta t} \quad (5)$$

popř.
$$= \int_{-\infty}^{\infty} \mathfrak{B}_\beta(n)e^{i\omega(n)\beta t}d\beta. \quad (6)$$

Kvantově teoreticky se zdá být nejjednodušší a nejpřirozenější, vztahy (3, 4) nahradit následujícími[n]:

$$\mathfrak{B}(n, n-\beta)e^{i\omega(n,n-\beta)t} = \sum_{\alpha=-\infty}^{\infty} \mathfrak{U}(n, n-\alpha)\mathfrak{U}(n-\alpha, n-\beta)e^{i\omega(n,n-\beta)t} \quad (7)$$

popř. $\quad = \int_{-\infty}^{\infty} d\alpha \; \mathfrak{U}(n, n-\alpha)\mathfrak{U}(n-\alpha, n-\beta)e^{i\omega(n,n-\beta)t};\quad (8)$

a sice dostáváme tento způsob skládání téměř nutně z kombinačních relací pro frekvence. Uděláme-li tento předpoklad (7) a (8) rozpoznáme také, že fáze kvantově teoretických $\mathfrak{U}$ mají stejně velký význam jako fáze v klasické teorii: jedině počátek času, a tedy fázová konstanta společná všem $\mathfrak{U}$, je libovolná a bez fyzikálního významu; fáze jednotlivých $\mathfrak{U}$ vstupuje[o] zásadně do veličiny $\mathfrak{B}$[1]. Geometrická interpretace takových kvantově teoretických fázových vztahů v analogii s klasickou teorií se zatím zdá sotva možná.

Ptáme-li se nyní na reprezentaci veličiny $x(t)^3$, nalezneme bez obtíží:

Klasicky:
$$\mathfrak{C}(n, \gamma) = \sum_{\alpha=-\infty}^{\infty}\sum_{\beta=-\infty}^{\infty} \mathfrak{U}_\alpha(n)\mathfrak{U}_\beta(n)\mathfrak{U}_{\gamma-\alpha-\beta}(n) \quad (9)$$

Kvantově teoreticky:
$$\mathfrak{C}(n, n-\gamma) = \sum_{\alpha=-\infty}^{\infty}\sum_{\beta=-\infty}^{\infty} \mathfrak{U}(n, n-\alpha)\mathfrak{U}(n-\alpha, n-\alpha-\beta)\mathfrak{U}(n-\alpha-\beta, n-\gamma) \quad (10)$$

popř. odpovídající integrály.

Podobným způsobem se nechají kvantově teoreticky vyjádřit všechny veličiny tvaru $x(t)^n$, a když je dána nějaká funkce $f[x(t)]$, můžeme k ní zřejmě vždy, když je tato funkce rozvinutelná do mocninné řady v $x$, nalézt kvantově teoretický analog. Pokud se však zabýváme dvěma veličinami $x(t)$, $y(t)$ a ptáme se na produkt $x(t)y(t)$ vyvstane závažná obtíž.

---

[1] Srov. Také H. A. Kramers a W. Heisenberg, *loc. cit.* Do tam použitých výrazů pro indukovaný elektrický moment vstupují fáze podstatným způsobem.

Nechť je $x(t)$ charakterizována pomocí $\mathfrak{U}$ a $y(t)$ pomocí $\mathfrak{B}$, pak dostáváme jako reprezentaci výrazu $x(t) \cdot y(t)$:

klasicky:
$$\mathfrak{C}_\beta(n) = \sum_{\alpha=-\infty}^{\infty} \mathfrak{U}_\alpha(n)\mathfrak{B}_{\beta-\alpha}(n).$$

Kvantově teoreticky:
$$\mathfrak{C}(n, n-\beta) = \sum_{\alpha=-\infty}^{\infty} \mathfrak{U}(n, n-\alpha)\mathfrak{B}(n-\alpha, n-\beta).$$

Zatímco klasicky bude vždy $x(t) \cdot y(t)$ rovno $y(t)x(t)$[p], nemusí to v kvantové teorii obecně platit. – Ve speciálních případech, např. při konstrukci výrazu $x(t) \cdot x(t)^2$, tato obtíž nevzniká.

Když se jedná, jako v případě otázky položené na začátku tohoto paragrafu, o konstrukce ve tvaru
$$v(t)\dot{v}(t),$$
budeme muset výraz $\dot{v}v$ kvantově teoreticky nahradit výrazem $\frac{v\dot{v}+\dot{v}v}{2}$, abychom dosáhli toho, že $v\dot{v}$ vystupuje jako derivace výrazu $\frac{v^2}{2}$.[q] Podobným způsobem se dají vždy uvést odpovídající kvantově teoretické střední hodnoty, které jsou hypoteticky v ještě vyšších řádech než rovnice (7) a (8).

Ohlédneme-li od právě popsaných obtíží mohly by rovnice typu (7), (8) obecně postačovat i k vyjádření vzájemného působení elektronů v atomu pomocí charakteristických amplitud elektronů.

§ 2. Po těchto úvahách, jejichž předmětem byla kinematika kvantové teorie, přejdeme k mechanickému problému, který směřuje k určení $\mathfrak{U}$, $\nu$, $W$ ze sil daných v systému. V dosavadní teorii se tento problém řeší ve dvou krocích:

1. Integrace pohybové rovnice
$$\ddot{x} + f(x) = 0. \tag{11}$$
2. Určení konstanty periodického pohyb vztahem
$$\oint p dq = \oint m\dot{x} dx = J(= nh). \tag{12}$$

Dáme-li si za úkol vybudovat kvantově teoretickou mechaniku, která je co možná nevíce analogická klasické, nabízí se převzít pohybovou rovnici (11) přímo do kvantové teorie, přičemž je pouze nutné – abychom nesešli z jistého základu principiálně

pozorovatelných veličin –, na místě veličin $\ddot{x}$, $f(x)$ nasadit jejich kvantově teoretické reprezentace známé z § 1. V klasické teorii je možné hledat řešení rovnice (11) vyjádřením $x$ ve formě Fourierových řad, popř. Fourierových integrálů s neznámými koeficienty (a frekvencemi); samozřejmě pak obecně obdržíme nekonečně mnoho rovnic s nekonečně mnoha neznámými, popř. integrální rovnice, které se jen ve speciálních případech dají přetransformovat do jednoduchých rekurzivních formulí pro $\mathfrak{U}$. V kvantové teorii jsme předběžně odkázáni na tento způsob řešení (11), neboť, jak jsme výše diskutovali, nelze definovat žádnou kvantově teoretickou funkci přímo analogickou funkci $x(n,t)$.

To má za následek, že kvantově teoretické řešení rovnice (11) je prozatím proveditelné pouze v těch nejjednodušších případech. Než přistoupíme k těmto jednoduchým příkladům, odvoďme ještě kvantově teoretické stanovení konstanty podle rovnice (12). Předpokládejme tedy, že je pohyb (klasicky) periodický[r]:

$$x = \sum_{\alpha=-\infty}^{\infty} a_\alpha(n) e^{i\alpha\omega_n t} ; \qquad (13)$$

pak je

$$m\dot{x} = m \sum_{\alpha=-\infty}^{\infty} a_\alpha(n) \cdot i\alpha\omega_n e^{i\alpha\omega_n t}$$

a

$$\oint m\dot{x}\,dx = \oint m\dot{x}^2\,dt = 2\pi m \sum_{\alpha=-\infty}^{\infty} a_\alpha(n) a_{-\alpha}(n) \alpha^2 \omega_n .$$

Jelikož dále platí $a_{-\alpha}(n) = \overline{a_\alpha(n)}$ ($x$ má být reálné) plyne

$$\oint m\dot{x}^2\,dx = 2\pi m \sum_{\alpha=-\infty}^{\infty} |a_\alpha(n)|^2 \alpha^2 \omega_n . \qquad (14)$$

Tento fázový integrál byl doposud většinou stavěn roven celočíselným násobkům $h$, tedy $n \cdot h$; taková podmínka nejen že zapadá jen velmi nuceně do mechanických výpočtů, ale ukazuje se i z dosavadního pohledu ve smyslu korespondenčního principu jako svévolná; neboť v rámci korespondence jsou $J$ určeny jako celočíselné násobky $h$ až na aditivní konstantu, a na místě rovnice (14) by mělo přirozeně stát:

$$\frac{d}{dn}(nh) = \frac{d}{dn} \cdot \oint m\dot{x}^2\,dt,$$

to znamená

$$h = 2\pi m \cdot \sum_{\alpha=-\infty}^{\infty} \alpha \frac{d}{dn}(\alpha\omega_n \cdot |a_\alpha|^2). \qquad (15)$$

Taková podmínka určuje ovšem koeficienty $a_\alpha$ také až na konstantu a tato neurčenost vedla empiricky k obtížím při výskytu poločíselných kvantových čísel.

Ptáme-li se na kvantově teoretický vztah mezi pozorovatelnými veličinami odpovídající rovnicím (14) a (15), chybějící jednoznačnost se sama znovu ustaví.

Sice lze právě jen rovnici (15) převést na jednoduchý kvantově teoretický tvar[s] spojený s K r a m e r s o v o u teorií disperze[1t]:

$$h = 4\pi m \sum_{\alpha=0}^{\infty} \{|a(n, n+\alpha)|^2 \omega(n, n+\alpha) - |a(n, n-\alpha)|^2 \omega(n, n-\alpha)\}, \quad (16)$$

přesto tento vztah zde postačí k jednoznačnému určení koeficientů $a$; neboť původně neurčená konstanta v koeficientech $a$ je určena podmínkou, že je dán nějaký základní stav[u], ze kterého již nenastává žádné záření; nechť je tento základní stav označen indexem $n_0$, pak mají být všechny

$$a(n_0, n_0 - \alpha) = 0 \ (\text{pro } \alpha > 0).$$

Otázka poločíselného nebo celočíselného kvantování nemůže tedy v kvantově teoretické mechanice, která používá pouze vztahy mezi pozorovatelnými veličinami, vyvstat.

Rovnice (11) a (16) dohromady zahrnují, pokud je možné je vyřešit, úplné určení nejen frekvencí a energií, ale také kvantově teoretických pravděpodobností přechodu. Skutečné matematické provedení se prozatím daří pouze v těch nejjednodušších případech: obzvláštní komplikace vznikají v mnoha systémech, jako např. u atomu vodíku, tím, že řešení odpovídá částečně periodickému a částečně aperiodickému pohybu, což má za následek, že kvantově teoretické řady (7), (8) a rovnice (16) se vždy rozpadají na jednu sumu a jeden integrál. Kvantově mechanicky se rozdělení na „periodický a aperiodický pohyb" obecně nedá provést.

Přesto bychom snad mohli, přinejmenším principiálně, považovat rovnice (11) a (16) za uspokojivé řešení mechanických problémů, pokud by bylo možné ukázat, že toto řešení souhlasí, popř. není v rozporu s dosud známými kvantově mechanickými vztahy; že tedy malá porucha mechanického problému vede na

---

[1] Tento vztah by již podán na v W. Kuhn, ZS. F. Phys. **33**, 408, 1925 a Thomas, Naturw. **13**, 1925 na základě studia disperze.

přídavné členy v energii, popř. ve frekvencích, které odpovídají právě výrazům nalezeným K r a m e r s e m a B o r n e m – v protikladu k těm, které by dávala klasická teorie. Dále musí být zjištěno, jestli také ve zde navrženém kvantově teoretickém přístupu odpovídá obecně rovnici (11) integrál energie $m\frac{\dot{x}^2}{2} + U(x) =$ const a jestli takto získaná energie – podobně jako platí klasicky: $\nu = \frac{\partial W}{\partial J}$ – odpovídá podmínce: $\Delta W = h \cdot \nu$. Teprve obecné zodpovězení těchto otázek by mohlo vysvětlit vnitřní souvislosti dosavadních kvantově mechanických výzkumů[v] a vést ke kvantové mechanice, která operuje důsledně jen s pozorovatelnými veličinami. S výjimkou obecného vztahu mezi K r a m e r s o v o u disperzní rovnicí a rovnicemi (11) a (16) můžeme výše položené otázky zodpovědět pouze ve velmi speciálních, jednoduchou rekurzí řešitelných případech.

Zmíněný obecný vztah mezi K r a m e r s o v o u disperzní teorií a našimi rovnicemi (11) a (16) spočívá v tom, že z rovnice (11) (tzn. její kvantově teoretické analogie) stejně tak jako v klasické teorii vyplývá, že se oscilující elektron vůči světlu, které má mnohem kratší vlnovou délku než všechny vlastní oscilace systému, chová jako volný elektron. Tento výsledek plyne také z K r a m e r s o v y teorie, vezme-li se navíc v úvahu rovnice (16). A skutečně nachází K r a m e r s pro moment[w] indukovaný vlnou $E \cos 2\pi \nu t$:

$$M = e^2 E \cos 2\pi \nu t \cdot \frac{2}{h} \sum_{\alpha=0}^{\infty} \left\{ \frac{|a(n, n+\alpha)|^2 \nu(n, n+\alpha)}{\nu^2(n, n+\alpha) - \nu^2} - \frac{|a(n, n-\alpha)|^2 \nu(n, n-\alpha)}{\nu^2(n, n-\alpha) - \nu^2} \right\}$$

tedy pro $\nu \gg \nu(n, n+\alpha)$

$$M = -\frac{2Ee^2 \cos 2\pi\nu t}{\nu^2 \cdot h} \sum_{\alpha=0}^{\infty} \{|a(n, n+\alpha)|^2 \nu(n, n+\alpha) - |a(n, n-\alpha)|^2 \nu(n, n-\alpha),$$

což díky (16) přechází na

$$M = -\frac{e^2 E \cos 2\pi \nu t}{\nu^2 \cdot 4\pi^2 m}.$$

§ 3. Jako nejjednodušší příklad bude v následujícím pojednán anharmonický oscilátor:

$$\ddot{x} + \omega_0^2 x + \lambda x^2 = 0. \qquad (17)$$

Klasicky lze tuto rovnici splnit předpokladem ve tvaru

$$x = \lambda a_0 + a_1 \cos \omega t + \lambda a_2 \cos 2\omega t + \lambda^2 a_3 \cos 3\omega t + \cdots \lambda^{\tau-1} a_\tau \cos \tau\omega t,$$

kde $a$ jsou mocninné řady v $\lambda$, které začínají členem nezávislým na $\lambda$. Kvantově teoreticky vyzkoušíme analogický předpoklad a reprezentujeme $x$ členy ve tvaru

$$\lambda a(n,n); \ a(n,n-1)\cos\omega(n,n-1)t; \ \lambda a(n,n-2)\cos\omega(n,n-2)t;$$
$$\ldots \lambda^{\tau-1} a(n,n-\tau)\cos\omega(n,n-\tau)t \ldots$$

Rekurzivní vztahy k určení koeficientů $a$ a $\omega$ znějí (až do členů řádu $\lambda$) podle rovnic (3), (4), popř. (7), (8):

Klasicky:

$$\left.\begin{aligned}
&\omega_0^2 a_0(n) + \frac{a_1^2(n)}{2} = 0; \\
&-\omega^2 + \omega_0^2 = 0; \\
&(-4\omega^2 + \omega_0^2)a_2(n) + \frac{a_1^2}{2} = 0; \\
&(-9\omega^2 + \omega_0^2)a_3(n) + a_1 a_3 = 0;
\end{aligned}\right\} \quad (18)$$

Kvantově teoreticky:

$$\left.\begin{aligned}
&\omega_0^2 a_0(n) + \frac{a^2(n+1,n) + a^2(n,n-1)}{4} = 0; \\
&-\omega^2(n,n-1) + \omega_0^2 = 0; \\
&(-\omega^2(n,n-2) + \omega_0^2)a(n,n-2) + \frac{a(n,n-1)a(n-1,n-2)}{2} = 0; \\
&(-\omega^2(n,n-3) + \omega_0^2)a(n,n-3) \\
&\quad + \frac{a(n,n-1)a(n-1,n-3)}{2} + \frac{a(n,n-2)a(n-2,n-3)}{2} = 0;
\end{aligned}\right\} \quad (19)$$

K tomu přicházejí kvantovací[x] podmínky:

Klasicky $(J = n\,h)$[y]:

$$1 = 2\pi m \frac{d}{dJ} \sum_{\tau=-\infty}^{\infty} \tau^2 \frac{|a_\tau|^2 \omega}{4}.$$

Kvantově teoreticky[z]:

$$h = \pi m \sum_{\tau=0}^{\infty} [|a(n+\tau,n)|^2 \omega(n+\tau,n) - |a(n,n-\tau)|^2 \omega(n,n-\tau)].$$

Toto dává v prvním přiblížení, jak klasicky, tak kvantově teoreticky:

$$a_1^2(n) \ \text{popř.} \ a^2(n,n-1) = \frac{(n+\text{const})h}{\pi m \omega_0}. \quad (20)$$

Kvantově teoreticky lze konstantu v rovnici (20) určit z podmínky, že $a(n_0, n_0 - 1)$ v základním stavu má být rovno nule. Budeme-li $n$ číslovat tak, že $n$ v základním stavu je rovno nule, tedy $n_0 = 0$, plyne

$$a^2(n, n-1) = \frac{n\,h}{\pi\,m\,\omega_0}.$$

Z rekurzivních rovnic (18) pak plyne, že v klasické teorii bude $a_\tau$ mít tvar $\kappa(\tau) n^{\frac{\tau}{2}}$, kde $\kappa(\tau)$ představuje faktor nezávislý na $n$. V kvantové teorii dostáváme z (19)

$$a(n, n-\tau) = \kappa(n) \sqrt{\frac{n!}{(n-\tau)!}}, \quad (21)$$

kde $\kappa(\tau)$ představuje stejný, na $n$ nezávislý proporcionální faktor. Pro velké hodnoty $n$ přecházejí samozřejmě kvantově teoretické hodnoty $a_\tau$ asymptoticky na klasické[aa].

Pro energii je nasnadě zkusit klasický předpoklad

$$\frac{m\dot{x}^2}{2} + m\omega_0^2 \frac{x^2}{2} + \frac{m\lambda}{3}x^3 = W,$$

který je skutečně ve zde propočítaném přiblížení i kvantově teoreticky konstantní a podle (19), (20) a (21) má hodnotu:

Klasicky:
$$W = \frac{n\,h\,\omega_0}{2\,\pi}. \quad (22)$$

Kvantově teoreticky [podle (7), (8)]:
$$W = \frac{\left(n + \frac{1}{2}\right)h\,\omega_0}{2\pi} \quad (23)$$

(až na členy řádu $\lambda^2$).

Podle tohoto přístupu tedy již u harmonického oscilátoru energie není reprezentovatelná „klasickou mechanikou", tzn. rovnicí (22), nýbrž má tvar rovnice (23).

Přesnější propočet včetně vyšších přiblížení ve $W, a, \omega$ bude proveden na jednoduchém příkladu anharmonického oscilátoru typu:

$$\ddot{x} + \omega_0^2 x + \lambda x^3 = 0.$$

Klasicky tu můžeme postavit:
$$x = a_1 \cos \omega t + \lambda a_3 \cos 3\omega t + \lambda^2 a_5 \cos 5\omega t + \cdots,$$

analogicky zkusíme kvantově teoreticky předpoklad:
$$a(n, n-1)\cos\omega(n, n-1)t\,;\ \lambda a(n, n-3)\cos\omega(n, n-3)t\,;\ \ldots$$

Veličiny $a$ jsou opět mocninné řady v $\lambda$, jejichž první člen má, stejně jako v rovnici (21), tvar:

$$a(n, n-\tau) = \kappa(\tau)\sqrt{\frac{n!}{(n-\tau)!}},$$

jak obdržíme propočtením rovnic odpovídajících rovnicím (18), (19).

Provedeme-li výpočet $\omega$, $a$ podle (18), (19) do přiblížení $\lambda^2$ popř. $\lambda$, obdržíme:

$$\omega(n, n-1) = \omega_0 + \lambda \cdot \frac{3\,n\,h}{8\pi\omega_0^2 m} - \lambda^2 \cdot \frac{3\,h^2}{256\,\omega_0^5 m^2 \pi^2}(17n^2 + 7) + \cdots \quad (24)$$

$$a(n, n-1) = \sqrt{\frac{n\,h}{\pi\omega_0 m}}\left(1 - \lambda\frac{3\,n\,h}{16\pi\omega_0^3 m} + \cdots\right). \quad (25)$$

$$a(n, n-3) = \frac{1}{32}\sqrt{\frac{h^3}{\pi^3\omega_0^7 m^3}n(n-1)(n-2)}\left(1 - \lambda\frac{39(n-1)h}{32\pi\omega_0^3 m}\right). \quad (26)$$

Energie, která je definovaná jako konstantní člen výrazu

$$m\frac{\dot{x}^2}{2} + m\omega_0^2\frac{x^2}{2} + \frac{m\lambda}{4}x^4$$

(že jsou všechny periodické členy skutečně nulové jsem nemohl obecně dokázat, v propočtených členech to taky bylo), dává

$$W = \frac{\left(n + \frac{1}{2}\right)h\omega_0}{2\pi} + \lambda \cdot \frac{3\left(n^2 + n + \frac{1}{2}\right)h^2}{8 \cdot 4\pi^2\omega_0^3 \cdot m}$$

$$-\lambda^2 \cdot \frac{h^3}{512\,\pi^3\omega_0^5 m^2}\left(17n^3 + \frac{51}{2}n^2 + \frac{59}{2}n + \frac{21}{2}\right). \quad (27)$$

Tuto energii můžeme také vypočítat ještě pomocí K r a m e r s o v a-B o h r o v a postupu, ve kterém se člen $\frac{m\lambda}{4}x^4$ vezme jako porucha k harmonickému oscilátoru. Dospějeme pak skutečně znovu přesně k výsledku (27), což mi připadá jako pozoruhodná opora pro kvantově mechanické rovnice, na nichž je tento výsledek založen. Energie spočítaná podle (27) dále splňuje rovnici [srov. (24)]:

$$\frac{\omega(n, n-1)}{2\pi} = \frac{1}{h} \cdot [W(n) - W(n-1)],$$

kterou rovněž musíme uvažovat jako nutnou podmínku možnosti určení přechodových pravděpodobností odpovídajících rovnicím (11) a (16).

Na závěr uveďme jako příklad tuhý rotátor[bb] a poukažme na vztahy rovnic (7), (8) k výrazům pro intenzity Zeemanova efektu[1] a multipletů[2].

Nechť je rotátor reprezentován elektronem, který krouží v k o n s t a n t n í vzdálenosti $a$ kolem jádra. „Pohybové rovnice" pak uvádějí jak klasicky, tak kvantově pouze, že elektron opisuje rovnoměrnou rotaci v konstantní vzdálenosti $a$ kolem jádra s úhlovou rychlostí $\omega$. Kvantovací podmínky (16) dávají podle (12):

$$h = \frac{d}{dn}(2\pi m a^2 \omega),$$

podle (16):

$$h = 2\pi m\{a^2 \omega(n+1, n) - a^2 \omega(n, n-1)\},$$

z čehož v obou případech plyne:

$$\omega(n, n-1) = \frac{h \cdot (n + \text{const})}{2\pi m a^2}.$$

Podmínka, že v základním stavu $(n_0 = 0)$ má záření vymizet, vede na vztah:

$$\omega(n, n-1) = \frac{h \cdot n}{2\pi m a^2}. \tag{28}$$

Energie bude

$$W = \frac{m}{2} v^2$$

nebo podle (7), (8)

$$W = \frac{m}{2} a^2 \cdot \frac{\omega^2(n, n-1) + \omega^2(n+1, n)}{2} = \frac{h^2}{8\pi^2 m a^2}\left(n^2 + n + \frac{1}{2}\right), \tag{29}$$

což opět splňuje vztah $\omega(n, n-1) = \frac{2\pi}{h}[W(n) - W(n-1)]$. Za oporu rovnicím (28) a (29), které se od dosud běžných teorií odchylují, můžeme považovat, že podle K r a t z e r a[3] se zdají mnohá pásová spektra (také ta, ve kterých je existence elektronové hybnosti nepravděpodobná) vyžadovat rovnice typu (28), (29) (které byly kvůli klasické mechanické teorii doposud vysvětlovány poločíselným kvantováním).

---

[1] Gouldsmit und R. de L. Kronig, Naturw. **13**, 90, 1925; H. Hönl, ZS. F. Phys. **31**, 340, 1925.

[2] R. de L. Kronig, ZS f. Phys. **31**, 885, 1925; A. Sommerfeld und H. Hönl, Sitzungsber. d. Preuß. Akad. d. Wiss. 1925, S. 141; H. N. Russell, Nature **115**, 835, 1925.

[3] Srov. např. B. A. Kratzer, Sitzungsber. d. Bayr. Akad. 1922, S. 107.

Abychom u rotátoru došli ke G o u d s m i t o v ý m-K r o n i g o v ý m-H ö n l o v ý m rovnicím, musíme opustit oblast problémů s jedním stupněm volnosti a předpokládat, že rotátor koná v nějakém směru v prostoru kolem osy $z$ nějakého vnějšího pole velmi pomalou precesi o. Nechť se kvantové číslo odpovídající této precesi značí $m$. Pak bude pohyb reprezentován veličinami

$$z:\ a(n,n-1;m,m)\cos\omega(n,n-1)t;$$
$$x+iy:\ b(n,n-1;m,m-1)e^{i[\omega(n,n-1)+o]t};$$
$$b(n,n-1;m-1,m)e^{i[-\omega(n,n-1)+o]t}.$$

Pohybové rovnice znějí jednoduše:
$$x^2+y^2+z^2=a^2,$$
což podle (7) vede k rovnicím[1]:

$$\frac{1}{2}\left\{\frac{1}{2}a^2(n,n-1;m,m)+b^2(n,n-1;m,m-1)+b^2(n,n-1;m,m+1)\right.$$
$$\left.+\frac{1}{2}a^2(n+1,n;m,m)+b^2(n+1,n;m-1,m)+b^2(n+1,n;m+1,m)\right\}=a^2. \quad (30)$$

$$\frac{1}{2}a(n,n-1;m,m)a(n-1,n-2;m,m)$$
$$=b(n,n-1;m,m+1)b(n-1,n-2;m+1,m)$$
$$+b(n,n-1;m,m-1)b(n-1,n-2;m-1,m). \quad (31)$$

K tomu patří podle (16) kvantovací podmínka:
$$2\pi m\{b^2(n,n-1;m,m-1)\omega(n,n-1)$$
$$-b^2(n,n-1;m-1,m)\omega(n,n-1)\}=(m+\text{const})h. \quad (32)$$

Klasické vztahy odpovídající těmto rovnicím:
$$\left.\begin{array}{c}\frac{1}{2}a_0^2+b_1^2+b_{-1}^2=a^2;\\ \frac{1}{4}a_0^2=b_1b_{-1};\\ 2\pi m(b_{+1}^2-b_{-1}^2)\omega=(m+\text{const})h\end{array}\right\} \quad (33)$$

postačují (až na neurčenou konstantu u $m$) k jednoznačnému určení $a_0$, $b_1$, $b_{-1}$.

Nejjednodušší řešení kvantově teoretických rovnic (30), (31), (32), které se nabízí, zní:

$$b(n,n-1;m,m-1)=a\sqrt{\frac{(n+m+1)(n+m)}{4\left(n+\frac{1}{2}\right)n}};$$

$$b(n,n-1;m-1,m)=a\sqrt{\frac{(n-m)(n-m+1)}{4\left(n+\frac{1}{2}\right)n}};$$

$$a(n,n-1;m,m)=a\sqrt{\frac{(n+m+1)(n-m)}{\left(n+\frac{1}{2}\right)n}}.$$

---

[1] Rovnice (30) je v podstatě identická s O r n s t e i n o v ý m i-B u r g e r o v ý m i sumačními pravidly.

Tyto výrazy souhlasí s rovnicemi G o u d s m i t a, K r o n i g a  a  H ö n l a; nelze však snadno nahlédnout, že tyto výrazy představují j e d i n é řešení rovnic (30), (31), (32) – což se mi ovšem při zohlednění okrajových podmínek (vymizení *a*, *b* na „okrajích", srov. výše citované práce K r o n i g a, S o m m e r f e l d a  a  H ö n l a, R u s s e l l a) zdá pravděpodobné.

Úvahy podobné těm, které jsme zde podnikli vedou také u vztahů pro intenzity multipletů k výsledku, že uvedená pravidla pro intenzity souhlasí s rovnicemi (7) a (16). Tyto výsledky je opět možné použít jako oporu zejména pro správnost kinematické rovnice (7).

Zda metodu určení kvantově teoretických dat pomocí vztahů mezi pozorovatelnými veličinami, jak je zde navržena, lze již v principu považovat za uspokojivou, nebo zda tato metoda představuje přece jen příliš hrubý pokus o řešení[cc] zatím zřejmě velmi spletitého fyzikálního problému kvantově teoretické mechaniky, se ukáže teprve hlubším matematickým zkoumáním této zde pouze povrchně použité metody.

G ö t t i n g e n, Ústav teoretické fyziky.



# Poznámky

[a] Autorem původně zamýšlený překlad zněl: „O kvantově teoretickém přehodnocení kinematických a mechanických vztahů". V souladu s existující literaturou byl název práce nakonec převzat z knihy Ivana Štolla, *Dějiny fyziky*, Prometheus, Praha 2009, str. 467.

[b] Přihlédnuto k anglickému překladu: B. L. van der Waerden, ed., *Sources of Quantum Mechanics*, Dover Publications, Inc., New York, 1968.

[c] Jde o zřejmé odmítnutí kinematiky, protože se vyžadují částicové vlastnosti světla, které nelze do vlnového popisu (kinematiky) žádným způsobem zahrnout.

[d] Dvě v kontextu překládané práce významově identická německá slova „Ort" a „Lage" tu překládáme také dvěma, pro účely tohoto textu identickými, českými výrazy „poloha" a „pozice".

[e] V originálu je použito slovo *Ansatz*, které má v matematice a fyzice svůj zúžený význam (česky bychom nejspíše řekli „předpoklad" nebo dokonce „nástřel"), se kterým se používá v nezměněné formě i v anglické vědecké literatuře. Je téměř jisté, že Heisenberg tu nemyslel tento konkrétní význam, spíše míní, že předchozí práce představují první kroky správným směrem. Je ale třeba uvážit, že výraz „Ansatz" má v kontextu fyziky svoje specifické konotace. Vzhledem k tomu, jak je ve zbytku článku používána Kramersova teorie disperze, vůči níž dokonce Heisenberg poměřuje správnost svých výsledků, lze o Kramersově teorii hovořit i jako o předpokladu Heisenbergovy práce.

[f] Citovaná práce „v tisku" v pozn. 2 na str.880 je M. Born und P. Jordan, ZS. F. Phys. **33**, 479, 1925

[g] Zde je hranatými závorkami značen známý vektorový součin, tedy $[r, \dot{v}] = r \times \dot{v}$. Frakturová písmena $\mathfrak{E}$ (velké E) a $\mathfrak{H}$ (velké H) standardně představují elektrické a magnetické pole v radiační (také vlnové) oblasti i v poměrně nedávných učebnicích (např. J. Kvasnica, *Teorie elektromagnetického pole*, Academia, Praha, 1983).

[h] V originále je na tomto místě souvětí s podmětem v první větě (místo „určující" by mělo být „a určují"). Přestože Heisenberg takto spojuje věty poměrně často, jsou v překladu příslušné věty většinou upraveny, protože doslovný český překlad není obvykle dobře srozumitelný.

[i] Protože souvisejí s intenzitou záření, která je dána kvadráty těchto veličin.

[j] V originále stojí „kommensurable". Česko-Německý slovník Fr. Št. Kotta ([online] [cit. 2.7.2021] dostupné z: http://kott.ujc.cas.cz/index.php?vstup=soum%EC%F8iteln%FD &idHeslo=307084&zpusob=heslo&hledat=vse&popis=&heslo=) uvádí: „V geometrii slují souměřitelnými čáry, plochy, tělesa a i úhly, jichž poměr velikosti lze vyjádřiti celými čísly". České slovo „souměřitelný", tak jak je používáno v geometrii, má tedy přesně ten správný požadovaný význam (více viz. následující pozn.).

[k] V klasickém periodickém pohybu se vyskytují kromě základní frekvence obecně také všechny její celočíselné násobky, tzv. vyšší harmonické (frekvence). U kvantových systémů se vyšší harmonické v tomto smyslu nevyskytují. Tedy, na rozdíl od klasického periodického pohybu, složíme-li dvě frekvence kvantového pohybu, získáme obecně frekvenci, která se v tomto systému nevyskytuje. Tento náhled se v dalším ukáže jako naprosto zásadním vodítkem pro hledání nové kinematiky vhodné pro kvantovou teorii.

[l] V rovnici (2a) a dalších jsme se uchýlili k mírné modernizovali notace. V originále se sumace přes index $\alpha$ (nebo jakýkoliv jiný index) označuje za znaménkem sumace, tedy $\sum \alpha$ (index je graficky vložen do velkého řeckého sigma). Původní notace je nejen špatně reprodukovatelná při sazbě moderními prostředky (např. MS Word), ale je i poněkud matoucí, neboť vypadá jako, že se celý výraz násobí $\alpha$. Proto byla notace změněna na moderní, kde se $\alpha$ píše pod znaménko sumace.

---

<sup>m</sup> Obvykle slovo „zřejmě" v češtině vyjadřuje nejistotu mluvčího; zde se ale jedná o jeho doslovný význam (mohli bychom psát i „očividně"), tak jak se toto slovo používá v matematické a fyzikální literatuře. Rovnice (3) je pro čtenáře zběhlého ve vyšší matematice zřejmá.

<sup>n</sup> Doslovně: „… zdá být nejjednodušším a nejpřirozenějším předpokladem …"

<sup>o</sup> Zde by mělo nejspíše stát „fáze jednotlivých $\mathfrak{U}$ vstupují …", držíme se však originálu, kde je jednotné číslo (jako u „fáze" tak u slovesa „vstupovat").

<sup>p</sup> V celém článku se místy nesystematicky používá tečka k označení součinu. Někde je jí třeba pro grafické odlišení víceznakových symbolů od násobení dvou veličin, ale v zásadě vždy platí, že, např. $a \cdot b = ab$ a tečku lze beztrestně vynechat, jako se to stalo zde.

<sup>q</sup> V Heisenbergově notaci ještě není dobře ustaveno rozlišení klasických veličin a jejich kvantově mechanických reprezentací. Ve větě „…budeme muset výraz $v\dot{v}$ kvantově teoreticky nahradit výrazem $\frac{v\dot{v}+\dot{v}v}{2}$, abychom dosáhli toho, že $v\dot{v}$ vystupuje jako derivace výrazu $\frac{v^2}{2}$" se míní, že kvantová reprezentace původního klasického výrazu $v\dot{v}$ má v odpovídajících rovnicích vystupovat jako derivace výrazu $\frac{v^2}{2}$". Při nekomutativitě $v$ a $\dot{v}$ dává derivace $v^2$ podle času právě $v\dot{v} + \dot{v}v$.

<sup>r</sup> V rovnicích (13) až (15) Heisenberg drobně mění notaci frekvencí. Místo $\omega_n$ by tu konzistentně mělo stát $\omega(n)$.

<sup>s</sup> Rovnice (16) obsahuje chybné pořadí indexů u prvního ze dvou $a$ a první ze dvou frekvencí $\omega$. Zatímco v důsledku absolutní hodnoty nemá tato chyba u $a$ žádný vliv na výsledek, vedla by chyba u frekvence na opačné znaménko celého prvního ze dvou výrazů pod sumou. Správně má rovnice (16) znít:

$$h = 4\pi m \sum_{\alpha=0}^{\infty} \{|a(n+\alpha,n)|^2 \omega(n+\alpha,n) - |a(n,n-\alpha)|^2 \omega(n,n-\alpha)\}.$$

Výraz je použit se stejnou chybou ještě dvakrát na str. 887, zatímco na str. 888, těsně před rovnicí (20), je již vypsán se správnými indexy u koeficientů. Zde ale zdánlivě chybí faktor 4, jak je podrobně vysvětleno v pozn. (z).

<sup>t</sup> Druhá citace z pozn. 1 na str. 886 je neúplná. Přesná citace zní: W. Thomas, Naturw. **13**, 627, 1925.

<sup>u</sup> „Normalzustand" neboli normální stav se v dnešní literatuře označuje jako základní stav.

<sup>v</sup> V originále zde stojí „Versuche", tedy „pokusů", ve smyslu „pokusů o kvantovou mechaniku". Při překladu jsme přihlédli k anglickému překladu: B. L. van der Waerden, ed., *Sources of Quantum Mechanics*, Dover Publications, Inc., New York, 1968.

<sup>w</sup> Následující rovnice obsahují stejnou chybu jako rovnice (16) – viz. pozn. (s).

<sup>x</sup> Lze překládat „kvantové" i „kvantovací". Druhý překlad podtrhuje fakt, že podmínky umožňují kvantování, nebo přesněji, jejich splnění přímo vede na kvantování.

<sup>y</sup> V této rovnici chybí indikace, přes který index se sčítá. Je to samozřejmě přes $\tau$, které se vyskytuje u koeficientu $a$, a které odpovídá indexu $\alpha$ v rovnici (14).

<sup>z</sup> V této rovnici chybí oproti rovnici (16) faktor 4, ale naopak se uvádí správné pořadí indexů u koeficientů $a$ a $\omega$. Ztráta faktoru 4 je důsledkem změny definice koeficientů $a_\tau$, kterou Heisenberg provedl. V první části článku jsou jako $a$ označeny koeficienty komplexních exponenciál, zatímco zde jde o koeficienty cosinů, v nichž vyskytuje faktor $\frac{1}{2}$ navíc. Ve druhé mocnině to pak dává faktor ¼. Obě rovnice před rovnicí (20) jsou tedy uvedeny správně, pouze ve změněné notaci.

---

aa Zde se myslí, že $a(n, n - \tau)$ (jakožto kvantově teoretický výraz pro $a_\tau$) přechází pro velká kvantová čísla $n$ na klasické hodnoty $a_\tau$.

bb V době publikace článku se jednalo o všeobecně známý a v rámci kvantové teorie studovaný modelový problém, a tak si ho může Heisenberg dovolit uvést označením s určitým členem, tedy „der Rotator".

cc V originále stojí „einen viel zu groben Angriff", tedy doslovně „příliš hrubý útok" na problém (vytvoření správné) kvantové mechaniky. „Grob - hrubý" zde má znamenat „s nedostatečným rozlišením", jako např. u digitálního obrazu, což české spojení se slovem „útok" neevokuje. Německé slovo „Angriff – útok" má stejný kořen jako slova „Griff - úchop", „greifen – vzít", „ergreifen – uchopit" atd.



\end{document}